\documentclass[12pt,epsf]{article}
\usepackage{graphicx, amsmath}

\begin{document}

\sloppy
\begin{flushright}{SIT-HEP/TM-38}
\end{flushright}
\vskip 1.5 truecm
\centerline{\large{\bf Hybrid Curvatons from Broken Symmetry}}
\vskip .75 truecm
\centerline{\bf Tomohiro Matsuda\footnote{matsuda@sit.ac.jp}}
\vskip .4 truecm
\centerline {\it Laboratory of Physics, Saitama Institute of Technology,}
\centerline {\it Fusaiji, Okabe-machi, Saitama 369-0293, 
Japan}
\vskip 1. truecm
\makeatletter
\@addtoreset{equation}{section}
\def\theequation{\thesection.\arabic{equation}}
\makeatother
\vskip 1. truecm

\begin{abstract}
\hspace*{\parindent}
We present a new general mechanism for generating curvature
 perturbations after inflation. 
Our model is based on the simple assumption that a field that starts to
 oscillate after inflation has a potential characterized by an
 underlying global symmetry that is slightly or badly broken. 
Inhomogeneous preheating occurs due to the oscillation with the broken
 symmetry. Unlike the traditional curvaton model, we will not identify
 the curvaton with the oscillating field. 
The curvaton is identified with the preheat field that could be either a
 scalar, vector, or fermionic field. We introduce an explicit mass term
 for the curvaton, which is important for later evolution and 
 the decay. 
Our present model represents the simplest example of the hybrid of the
 curvatons and inhomogeneous preheating. 
\end{abstract}

\newpage
\section{Introduction}
\hspace*{\parindent}
The traditional inflationary scenario is based on the assumption that
the primordial curvature perturbation is generated by the fluctuation of
the inflaton field, which leads to constraints on the traditional
inflaton potential, as it is responsible for both the sufficient amount
of expansion of the Universe and the generation of the curvature
perturbations. 
Although this traditional scenario looks very simple and perhaps quite
convincing, the traditional scenario sometimes requires serious
fine-tunings, especially in low-scale inflationary
models\cite{low_inflation}. 
For example, the spectrum of the primordial curvature perturbation for
the traditional scenario is given by \cite{EU-book}
\begin{equation}
{\cal P}_{\cal R}(k)=\frac{1}{24\pi^2 M_p^4}\frac{V_I}{\epsilon_I},
\end{equation}
where $\epsilon_I$ denotes the conventional slow-roll parameter.
Obviously, fine tuning for the slow-roll parameter $\epsilon_I$ 
will be required for small inflationary scale.
Since low-scale inflation will be very important if some gravitational
effect is observed in LHC, alternatives to the traditional scenario may
be desired in the near future\footnote{Of course there are many other
problems in low-scale inflation that are related to other cosmological
issues. We think the most significant condition appears from
baryogenesis\cite{low-baryo, low-AD, low_DG} in case the electro-weak
baryogenesis fails.} 
Recently, many alternatives to the traditional inflation have been
discussed by many authors\cite{curvaton_1, matsuda_curvaton,
alternate, alt2, Inho_Reh_Dvali, alternate2, SSB-inst, 
matsuda_inst1, matsuda_inst2}.\footnote{The original curvaton
is not suitable for low-scale inflation because of the strict lower
bound for the Hubble parameter\cite{curvaton_low}. A
solution to this problem is discussed by us in the first reference in
Ref. \cite{matsuda_curvaton}. Hybrid curvaton is an another solution to
this problem.} 
In these alternatives the generation of the curvature perturbation is
induced by the late-time conversion mechanism that characterizes the
scenario. 
In these models both the ``seed'' isocurvature fluctuation is generated
along with exit horizon during primordial inflation.
Then the seed fluctuation is converted into curvature perturbation by
the late-time mechanism.
In these alternatives, and since the generation and the horizon exit of
the seed fluctuations occurs during the primordial inflation, the
typical length-scale is very large even though the conversion occurs
very late in the scenario. 

Based on recent progress in these ``alternatives'', in the present paper
we will consider the case where the oscillation that is associated with
preheating has a potential characterized by broken global symmetry. 
In Sec.2 we will discuss inhomogeneous preheating with slightly broken
symmetry.\footnote{One other important idea called ``inhomogeneous
reheating'' has been discussed by many authors\cite{Inho_Reh_Dvali,
alt2, alternate2}. Note that preheating is supposed to occur prior to
reheating.} 
The oscillating field may or may not be related to the chaotic-type
inflaton field.  
Inhomogeneous preheating with the same potential has been
studied in Ref. \cite{SSB-inst} from a different viewpoint.
Namely, the authors of Ref. \cite{SSB-inst} assumed that the preheat
field decays instantly into light fields.\footnote{Although the instant
decay is not considered in this paper, it will be important to explain
how this mechanism works in Ref. \cite{SSB-inst} and why we consider
curvatons\cite{curvaton_1, matsuda_curvaton, topological_curv} with the
same potential. See the appendix for more details and the constraints
that appear for the instant-decay scenario.
Note also that the authors of Ref. \cite{SSB-inst} assumed slightly broken
symmetry($x\ll 1$).}
The potential with broken
global symmetry would be very natural, and also preheating would be very
common if a field with an interaction starts to oscillate after
inflation.
Since our aim in this paper is to propose a simple and natural
realization of the hybrid of the curvatons and inhomogeneous preheating,
we will consider hybrid curvatons that work  
with the oscillating field accompanied by preheating, and whose
potential is characterized by broken global symmetry. 
We will also consider late-time oscillation that induces production of
the hybrid curvatons. 
Some variants will be discussed in the last section.  
 
In the original scenario of the curvatons, an oscillating field 
(this field is identified with the curvaton in the original curvaton
model) 
starts to dominate the energy density of the Universe because the ratio
of the oscillating field to the total energy density increases during
the radiation-dominated era. 
Then the curvatons decay to generate the cosmological perturbations. 
In the present paper we will consider a hybrid version of the original
curvaton model, in which (unlike the traditional curvaton model) the
curvaton is ``not'' the oscillating field. 
Instead of considering curvaton oscillation, we will consider generation
of massive curvatons through preheating.
In order to generate the initial perturbation of the curvaton number
density, we will consider inhomogeneous preheating induced by a
potential characterized by broken symmetry.
In this case the ``light field'' that is associated with the primordial
isocurvature perturbation is identified with the equipotential surface
of the potential, which is of course not in any sense the curvatons. 
One might think that this scenario is not the curvatons from the
viewpoint of the original discussions. 
However, we have already considered the basic idea in
Ref. \cite{topological_curv} where the fluctuating nucleation rate
induces the fluctuation of the number density of the ``topological
curvatons''.
We have also discussed another variation in Ref. \cite{NO_Curvatons}. 
We hope there is no further confusion in the name.

Summarizing the difference between the traditional curvaton model and
the hybrid version, the crucial difference is that in the original
curvaton scenario the primordial isocurvature perturbation is generated
for the scalar-field curvatons, while in the hybrid version the
primordial perturbation does not appear for the curvatons. 
In the hybrid model, the fluctuation related to other light scalar
fields\footnote{This field is identified with the $\theta$-direction in 
the model with slightly broken symmetry.}
is converted into the fluctuation of the curvaton number density after
inflation. 
The conversion is induced by the preheating, and the preheat field is
the curvaton, which could be a scalar, vector or fermionic field.

We will define a dimensionless parameter $\epsilon$ which measures the
ratio between the energy stored in the oscillating field and the
total energy density of the Universe.
This parameter is almost unity if the oscillating field is the
chaotic-type inflaton field, while it will be very small if one
considers late-time oscillation. 
What we will consider in this paper is the oscillation and preheating
with $0< \epsilon < 1$. 
$\epsilon \sim 1$ corresponds to the preheating induced by the
oscillation of the chaotic-type inflaton, while $\epsilon \ll 1$ is
possible if the oscillating field is not associated with inflaton.

\section{Slightly broken symmetry ($x\ll 1$)}
The basic idea for generating the fluctuation of the preheat-field
number density is not so different from the one discussed in
Ref. \cite{SSB-inst}, which is given in the appendix of this
paper. 
However, as we are considering long-lived curvatons as the preheat
field, there is an obvious discrepancy between the instant decay 
scenario\cite{SSB-inst, matsuda_inst1, matsuda_inst2} and the hybrid
curvatons. 
The potential we will consider for the hybrid curvatons is
\begin{equation}
V(\phi_1, \phi_2) = \frac{m^2}{2}\left[ \phi_1^2 + 
\frac{\phi_2^2}{1+x}\right],
\end{equation}
where the dimensionless parameter $x$ is a measure of the symmetry
breaking. 
Due to the oscillation of the scalar field $\phi\equiv \phi_1+ i\phi_2$,
the curvaton $\chi$ is produced at the ESP through preheating if there
is an interaction given by ${\cal L}= -\frac{1}{2}g^2 |\phi|^2
\chi^2$. 
In the instant-decay scenario\cite{SSB-inst}, the explicit mass term of
the preheat field was not important since the preheat field decays
instantly when $\phi$ reaches the critical value, where the 
effective mass of the preheat field is dominated by the huge value of
$\phi$. 
However, in the present scenario we will not assume an instant-decay
scenario. 
Instead, we will introduce an explicit mass term ${\cal L}=
-\frac{1}{2}(m_\chi^{bare})^2 \chi^2$ for the curvaton $\chi$, which becomes
important when we discuss the late-time decay of the curvaton. 
Then the comoving number density of the preheat field $\chi$ produced at 
the first scattering is given by
\begin{equation}
n_\chi^{(1)} = \frac{(g|\dot{\phi}_*|)^{3/2}}{(2\pi)^3}\exp \left[
-\frac{\pi g (m_\chi^2+|\phi_*|^2)}{|\dot{\phi}_*|}\right],
\end{equation}
where $*$ denotes the minimum of the potential along its trajectory.
The explicit form of $\phi_*$ and $\dot{\phi}_*$ are given in the
appendix. 
We will consider the primordial perturbation given by $\delta \theta_0$,
which leads to the fluctuation
\begin{eqnarray}
\frac{\delta |\phi_*|}{|\phi_*|} &=& \frac{2\cos 2\theta_0}{\sin 2\theta_0}
 \delta \theta_0 -(\delta \theta_0)^2\\
\frac{\delta |\dot{\phi}_*|}{|\dot{\phi}_*|} &=&
-x\left[A\delta \theta_0 + B (\delta \theta_0)^2\right],
\end{eqnarray}
where $A, B$ are defined by
\begin{eqnarray}
A &\equiv & \frac{\sin 2\theta_0}{2(1-x\sin^2 \theta_0)} \simeq 
\frac{\sin 2\theta_0}{2}+{\cal O}(x)\\
B &\equiv & \frac{4(1-x\sin^2 \theta_0)\cos 2\theta_0
+ x\sin^2 2\theta_0}{8(1-x\sin^2 \theta_0)^2} \simeq 
\frac{\cos 2\theta_0}{2}+{\cal O}(x).
\end{eqnarray}
Then the fluctuation of the preheat-field number density is 
\begin{eqnarray}
\label{deln_chi}
\frac{\delta n_\chi}{n_\chi} &\simeq &
\left(\frac{3}{2}+\frac{\pi g |\phi_*|^2}{|\dot{\phi}_*|}\right)
\frac{\delta|\dot{\phi}_*|}{|\dot{\phi}_*|}
-\frac{2\pi g |\phi_*|^2}{|\dot{\phi}_*|} \frac{\delta |\phi_*|}{\phi_*}\nonumber\\
&\simeq& 
-x\left(\frac{3}{2}+{\cal O}(x^2)\right)
\left[A\delta \theta_0 + B (\delta \theta_0)^2\right]
-\left({\cal O}(x^2)\right)\left[\frac{2\cos 2\theta_0}{\sin 2\theta_0}
 \delta \theta_0 -(\delta \theta_0)^2\right]\nonumber\\
&\simeq& -\frac{3x}{2}
\left[A\delta \theta_0 + B (\delta \theta_0)^2\right] + {\cal O}(x^2).
\end{eqnarray}

In the instant-decay scenario, the preheat field decays instantaneously
when $\phi$ reaches its maximum value just after the first scattering. 
Then, assuming that the cosmological perturbation is generated by the
instant decay, the decay products must nearly dominate the energy of the
Universe. 
This requirement puts a crucial bound on the dimensionless parameter
$\epsilon$.  
Since the energy density of the preheat field is bounded above by the
energy of the oscillating field, the requirement for the
preheat-field domination leads to $\epsilon \simeq 1$.
The condition $\epsilon \simeq 1$ suggests that the oscillating field
must be the inflaton of chaotic inflation, which finally leads to the
bounds that we have obtained in the appendix. 
On the other hand, $\epsilon \ll 1$ is acceptable for the curvatons
since the ratio of the curvaton to the total energy density will grow
during the radiation-dominated era. 

In the hybrid curvaton model, the condition from the COBE normalization
is given by 
\begin{equation}
\label{curv-x}
\left[\frac{3x \sin 2\theta_0}{4}\right]\delta \theta_0 \sim 10^{-5},
\end{equation}
which comes from Eq.(\ref{deln_chi}).
Here we will introduce a new parameter $\phi_{osc}$ which denotes the
initial value of the oscillating field at the time when the
oscillation starts.
Note that $\phi_{osc}\sim M_p$ in the instant-decay scenario
while 
$\phi_{osc} \ll M_p$ is allowed in the curvaton scenario.
The fluctuation $\delta \theta_0$ is determined by the value $\phi_I$
during horizon exit and the Hubble parameter $H_I$ during inflation, and
is given by $\delta \theta_0 \simeq H_I/2\pi\phi_I$.
This value becomes $\delta \theta_0 \sim H_I/20\pi M_p$ in the
instant-decay scenario.
On the other hand, in the curvaton scenario it is possible to have either
$\phi_I \gg \phi_{osc}$ or $\phi_I \ll \phi_{osc}$ since $\phi$ is not
necessarily the inflaton.
Moreover, $\delta \theta_0$ may evolve after the inflationary stage
until the time when the curvaton is generated through preheating. 
Hence, the fluctuation $\delta \theta_0$  is not a parameter that is
completely determined by the inflationary model. 

As we are considering long-lived curvatons, we cannot disregard the
confining potential that is induced as the backreaction of the preheat
field, and also the fluctuations that may be induced at the second and
the third scatterings. 
After the first scattering that we have discussed above, the scatterings
occur successively where the number densities are given by 
\begin{equation}
n_k^{j+1} =  b_k^j n_k^j.
\end{equation}
Here the coefficient $b$ is given by
\begin{equation}
b_k^j =1+2e^{-\pi \mu^2}-2\sin \theta^j e^{-\pi \mu^2/2}\sqrt{
1+e^{-\pi \mu^2}},
\end{equation}
where $\mu^2=(k^2+(m_\chi^{eff})^2)/g|\dot{\phi}|$ and $\theta^j$ is a
relative phase. 
Of course we cannot disregard the fluctuations coming from 
$\delta b_k^j$, although it will be very difficult to say which
scattering induces the most significant effect on the fluctuation. 
In Fig.1 we show an obvious example adding an effective mass
$m_{\chi}^{eff} = m^{bare}_\chi + g\phi$, where it will be 
 very difficult to examine the generic properties of the later
 scatterings. 
\begin{figure}[ht]
 \begin{center}
\begin{picture}(550,320)(0,0)
\resizebox{16cm}{!}{\includegraphics{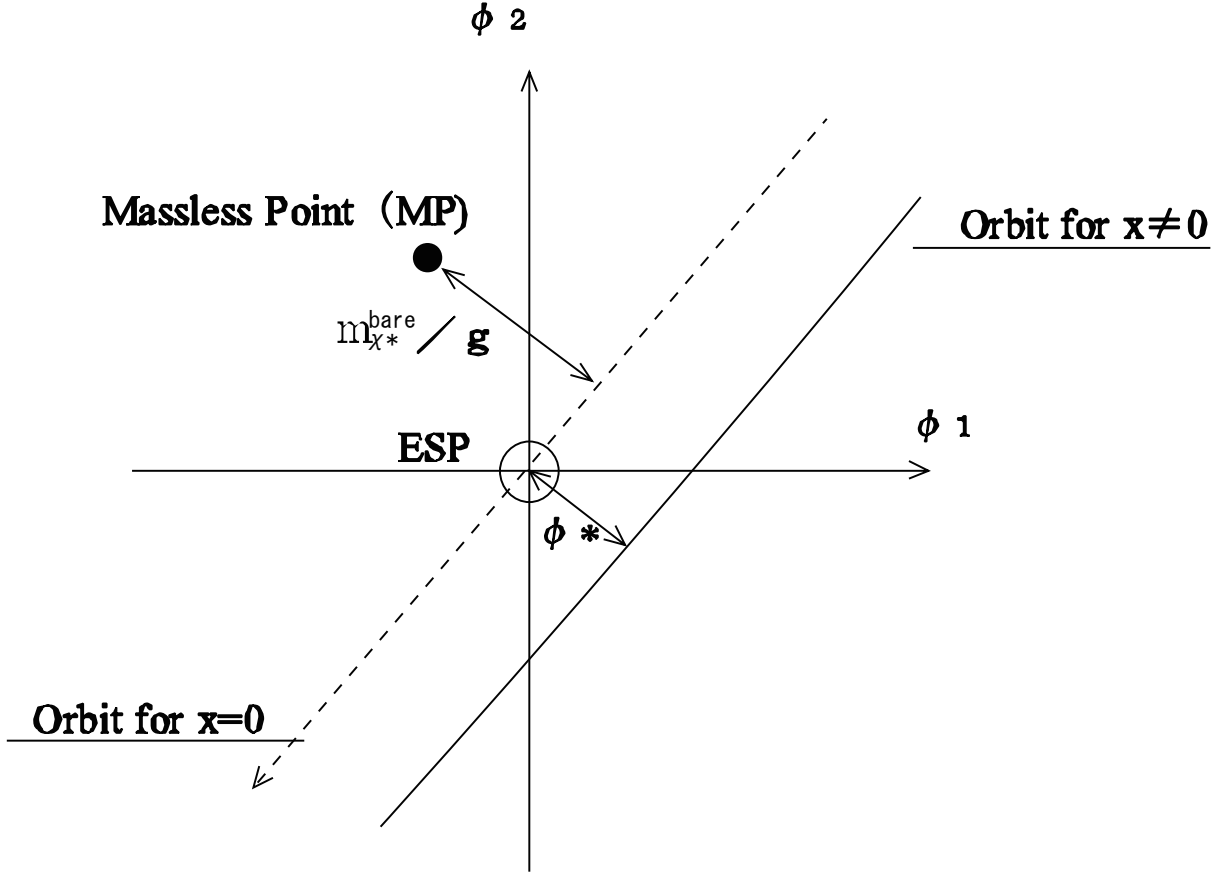}} 
\end{picture}
\caption{Including $m^{bare}_\chi \ne 0$, the massless point (MP) where preheat
  field becomes massless is shifted from the ESP.
Because of the shift denoted by $m_{\chi*}^{bare}$, the minimum value of
  the 
  preheat-field mass is given by $|m_\chi|=g\phi_{*} + m_{\chi*}^{bare}$.}
\label{fig1}
 \end{center}
\end{figure}
Instead of solving the process that is not essential for the
argument, we will simply assume that the first
scattering gives the dominant contribution to the fluctuation.
Note that even if the later scattering happens to generate a comparable
magnitude of fluctuation, the basic form of the result will not be
altered very much. 

Initially the curvaton energy density is 
$\rho_c =m_\chi n_\chi^{ini} \simeq (g\phi) (g
\dot{\phi}_*)^{3/2}/(2\pi)^3 \le (\dot{\phi}_*)^{2}\simeq m^2 \phi_{osc}^2$, 
while the 
total energy density of the Universe is $\rho_{tot} \simeq H_{osc}^2
M_p^2 \simeq m^2 M_p^2$, thus we obtain the initial ratio of the
energy density $r\equiv \rho_c/\rho_{tot}\le \phi_{osc}^2/M_p^2$,
which equation is similar to that found in the conventional curvatons.
The late-time evolution of the dimensionless parameter $r$ is also the
same as the conventional curvatons, but $\rho_\chi$ evolves as
$\rho_\chi\propto a^{-4}$\cite{dimo-trap} until $m^{bare}_\chi$ dominates
$m_\chi$. 
The only and the crucial difference appears from the condition coming
from the observed curvature perturbation.
In the conventional curvatons, this condition is given by
\begin{equation}
{\cal P}_\zeta \simeq\frac{rH_I}{2\pi \sigma_{osc}}\sim
10^{-5},
\end{equation}
where $\sigma_{osc}$ denotes the value of the curvaton $\sigma$ at the
onset of the oscillation.
This condition does not appear in our hybrid-type curvaton model, since
in our model the fluctuation of the curvaton number density is ``not''
related to the primordial isocurvature perturbation that might be
generated for the curvaton $\chi$, but is related to the parameter
$\delta \theta_0$ of the oscillating field $\phi$.
Of course the oscillating field $\phi$ is not the curvaton. 
In the conventional curvaton model, the decay rate of the curvaton is
supposed to be not much smaller than $\Gamma \sim m^3/M_p$, thus the
mass of the oscillating potential ($m$) puts a lower bound on the
decay temperature $T_d$. 
The bound from the decay temperature is of course important in the
present model.  
However, unlike the conventional curvatons, the decay rate of the
hybrid-type curvaton is not determined by the mass $m$ of the
oscillating field, but is determined by $m_\chi$ and it's coupling
$g_d$ to light field.
Therefore, in the hybrid-type curvaton the mass of the oscillating 
field($m$) that determines the initial value of the total energy density
through $H_{osc}\simeq m$ is not relevant to the decay temperature $T_d$,
which removes one of the most important conditions in the
conventional curvatons.

In the hybrid-type scenario, difficulty may appear from the condition
related to the dimensionless parameter $x$.
From Eq.(\ref{succ_pre}) in the appendix, the condition needed for
successful preheating is
\begin{equation}
\label{eq1}
x^2 < \frac{m}{g \phi_{osc}\sin^2 2\theta_0}.
\end{equation}
From Eq.(\ref{curv-x}), we can see the condition for the curvature
perturbation
\begin{equation}
\label{eq2}
x\sim \frac{4}{3 \sin 2\theta_0 } \frac{10^{-5}}{\delta \theta_0}.
\end{equation}
Finally, the non-Gaussian condition $f_{NL}<100$ is given by
\begin{equation}
\label{eq3}
x >\frac{1}{100} \left(\frac{\cos 2\theta_0}{ \sin^2 2\theta_0}\right).
\end{equation}
From Eq.(\ref{eq1}) and Eq.(\ref{eq2}), we obtain
\begin{equation}
\label{eq4}
\delta \theta_0 > 10^{-5}\sqrt{\frac{g \phi_{osc}}{m}}.
\end{equation}
From Eq.(\ref{eq3}) and Eq.(\ref{eq2}) we obtain
\begin{equation}
\label{eq5}
\delta \theta_0 < \frac{\sin 2\theta_0}{\cos 2\theta_0} 10^{-3}.
\end{equation}
Combining the last condition and Eq.(\ref{eq4}), we find the condition
for the ratio $\phi_{osc}/m$, which is given by
\begin{equation}
\label{eq6}
\frac{\phi_{osc}}{m}< 
\left(\frac{\sin 2\theta_0}{\cos 2\theta_0}\right)^2 \frac{10^4}{g}.
\end{equation}
Besides the above conditions, the ratio $r$ at the decay is bounded from
above according to
\begin{equation}
r(T_d) < \frac{\sqrt{m M_p}}{T_d}
\frac{m_\chi (g m\phi_{osc})^{3/2}}{m^2 M_p^2(2\pi)^3}< \frac{\sqrt{m M_p}}{T_d}
\frac{\phi_{osc}^2}{M_p^2},
\end{equation}
which is the usual condition that appears for the conventional curvatons.
Since the curvaton decay rate $\Gamma_\chi$ will be at least of order
$m_\chi^3/M_p^2$, there is a lower bound for $T_d$ given by
\begin{equation}
T_d \sim \sqrt{\Gamma M_p} \ge   \frac{m_{\chi}^{3/2}}{M_p^{1/2}}.
\end{equation}
These conditions lead to
\begin{equation}
r(T_d) < 10^{-2}\times
\frac{g^{3/2}\phi_{osc}^{3/2}}{M_p \sqrt{m^{bare}_\chi}} 
\simeq 10^{4}\times \frac{m^{3/2}}{\sqrt{m^{bare}_\chi} M_p}
\left|\frac{\sin 2\theta_0}{\cos 2\theta_0}\right|^3,
\end{equation}
where we assumed that the curvaton decays when $m_\chi^{bare}$ dominates
$m_\chi$. 
Therefore, our conclusion in this section is that one can expect $r(T_d)
\sim 1$ and the generation of the curvature perturbation from
the hybrid-type curvaton.\footnote{Alternatively, 
one may consider that the model helps to generate
smaller-scale perturbations that could be highly non-Gaussian.
This is an another possibility but it will not be discussed in this
paper.} 

In the next section we would like to see what happens in the opposite
limit of the broken symmetry, which is denoted by $x\gg 1$.

\section{Badly broken symmetry ($x\gg 1$)}
In the previous section we considered the hybrid-type curvaton that is
generated through inhomogeneous preheating with the potential with
$x \ll 1$. 
Alternatively, one may consider the opposite limit $x\gg 1$, 
where the scalar $\phi_2$ becomes very light compared with $\phi_1$.
Then, the oscillation will be induced solely by the ``massive'' field
$\phi_1$, while the fluctuation is dominantly induced by the ``light''
field $\phi_2$. 
In this limit, the two roles (oscillation and fluctuation) of the field
$\phi$ are distributed to the components
$\phi_1$ and $\phi_2$, which should be considered as independent
fields. 
In this section the value of the dimensionless parameter is supposed to
be about $x \sim 100$.\footnote{Note that an extremely light scalar
field will induce another cosmological problem to the theory.}

In the limit $x \gg 1$, we cannot use the ${\cal O}(x)$
expansion that was used in Eq.(\ref{deln_chi}).
In this limit the spectrum of the perturbations is given by $H_I/2\pi$
for both $\delta \phi_1$ and $\delta \phi_2$, while $\delta
\dot{\phi}_1$ and 
$\delta \dot{\phi}_2$ are given by $m H_I/2\pi$  and $0$, respectively. 
Then at the minimum of the trajectory, we obtain
$|\dot{\phi}_*|\simeq m \phi_{1,osc}$ and $|\phi_*| \simeq \phi_{2,osc}$.
Therefore, the fluctuation of the number density is given by
\begin{eqnarray}
\frac{\delta n_\chi}{n_\chi} &\simeq &
\left(\frac{3}{2}+\frac{\pi g |\phi_*|^2}{|\dot{\phi}_*|}\right)
\frac{\delta \dot{\phi}_*}{|\dot{\phi}_*|}
-\frac{2\pi g |\phi_*|^2}{|\dot{\phi}_*|} \frac{\delta \phi_*}{|\phi_*|}\nonumber\\
&\simeq &
\left(\frac{3}{2}+\frac{\pi g |\phi_{2,osc}|^2}{|m\phi_{1,osc}|}\right)
\frac{2\delta \phi_1}{|\phi_{1,osc}|} 
-\frac{2\pi g |\phi_{2,osc}|^2}{|m\phi_{1,osc}|} 
\frac{\delta \phi_{2}}{|\phi_{2,osc}|}
\nonumber\\
&\simeq& 
\frac{3\delta \phi_1}{|\phi_{1,osc}|} 
-\frac{2\pi g |\phi_{2,osc}|\delta \phi_{2}}{m |\phi_{1,osc}|},
\end{eqnarray}
where we assumed $\phi_{2,osc} \ll \phi_{1,osc}$.
We will disregard the first term assuming that $m$ is comparably smaller
than $|\phi_{2,osc}|$.
Then, the required condition for the COBE normalization is
\begin{equation}
\label{lxeq1}
\frac{g  H_I}{m}\left|\frac{\phi_{2,osc}}{\phi_{1,osc}}\right|\sim 10^{-5}.
\end{equation}
Note that in the scenario with slightly broken symmetry, the second term
that is proportional to $\delta \phi_*$ was negligible; while in the
present case the second term dominates the fluctuation.
Because of the exponential factor in the form of $n_\chi$, the required
condition for the successful preheating is 
\begin{equation}
\label{lxeq2}
\frac{\pi g |\phi_{2,osc}|^2}{m |\phi_{1,osc}|} <1.
\end{equation}
From Eq.(\ref{lxeq1}) and (\ref{lxeq2}), we find
\begin{equation}
\frac{\pi |\phi_{2,osc}|}{H_I}< 10^5.
\end{equation}
The non-Gaussian parameter $f_{NL}$ is of course different from the
$x\ll 1$ case, since in the present case the term proportional to
$\delta \phi_2$ will dominate the fluctuation.
Since the expression for $\delta n_\chi / n_\chi$ up to the second order
is given by
\begin{equation}
\label{2ndorder}
\frac{\delta n_\chi}{n_\chi} \simeq
-\frac{2\pi g |\phi_{2,osc}| \delta \phi_2}{|\dot{\phi}_*|} 
-\frac{1}{2}\left(\frac{2\pi g}{|\dot{\phi}_*|}
-\frac{4\pi^2 g^2 |\phi_{2,osc}|^2}{|\dot{\phi}_*|^2}
\right)(\delta \phi_{2,osc})^2,
\end{equation}
the non-Gaussian parameter is
\begin{equation}
-\frac{3}{5}f_{NL} \simeq
\frac{3|\dot{\phi}_*|}{4\pi g |\phi_{2,osc}|^2} 
-\frac{3}{2} \simeq \frac{3 m |\phi_{1,osc}|}{4\pi g |\phi_{2,osc}|^2}.
\end{equation}
Therefore, the condition $|f_{NL}|<100$ is expressed as
\begin{equation}
\frac{m|\phi_{1,osc}|}{\pi g |\phi_{2,osc}|^2} <10^2.
\end{equation}
Combined with Eq.(\ref{lxeq1}), we obtain
\begin{equation}
10^3 < \frac{\pi |\phi_{2,osc}|}{H_I}.
\end{equation}
As a result, we find the initial condition for the light field $\phi_2$ as
\begin{equation}
10^3 < \frac{\pi |\phi_{2,osc}|}{H_I}<10^{5}.
\end{equation}
As in the slightly broken case, the ratio $r$ at the decay is bounded
from above by 
\begin{equation}
r(T_d) < 10^{-2}\times
\frac{g^{3/2}\phi_{1,osc}^{3/2}}{M_p \sqrt{m^{bare}_\chi}} 
\end{equation}
which leads to the initial condition 
$(g \phi_{1,osc})^3 >  10^{4}M_p^2 m^{bare}_\chi$.
Therefore, generation of the curvature perturbation from the hybrid-type 
curvaton is possible for $x \gg 1$.

\section{Conclusions and Discussions}
\hspace*{\parindent}
In this paper we have considered inhomogeneous preheating that is
characterized by a potential with broken symmetry, and looked to see if
this could be used to generate curvatons. 
As we have shown in this paper, this simple realization of the hybrid of
the two seemingly different ideas is successful. 

To complement the discussion, we will summarize the situation in this
paper and compare it with other extended models. 
First, there is the possibility that the preheat field decays
instantaneously after preheating. 
This is the so-called instant preheating scenario\cite{instant-original}
which has been used to generate the cosmological perturbations in
 Ref. \cite{SSB-inst} for $x\ll 1$, and in Ref. \cite{matsuda_inst1, matsuda_inst2} for $x \gg
 1$.
Instead of considering the instant decay, we considered in this paper a
long-lived curvaton generation. 
Combined with the curvatons, the inhomogeneous preheating scenario opens
another possibility for generating the cosmological perturbations from
late-time oscillation.  

We used two scalar fields $\phi_1$ and $\phi_2$.
Our model is characterized by the $\phi_2$-potential, which is shown in
Fig.2 and Fig.3.
\begin{figure}[ht]
 \begin{center}
\begin{picture}(550,120)(0,0)
\resizebox{15cm}{!}{\includegraphics{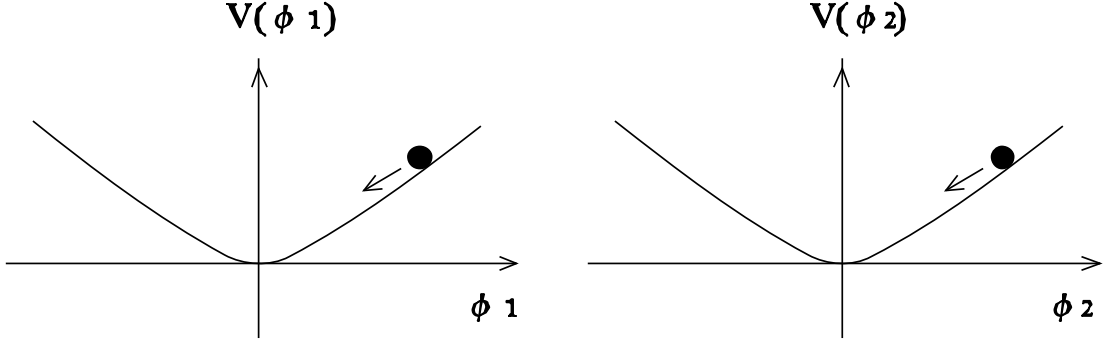}} 
\end{picture}
\caption{This figure shows the potential for slightly broken symmetry
  ($x\ll 1$). $\phi_1$ and $\phi_2$ start to oscillate at the same time.}
\label{fig2}
 \end{center}
\end{figure}
\begin{figure}[ht]
 \begin{center}
\begin{picture}(550,110)(0,0)
\resizebox{15cm}{!}{\includegraphics{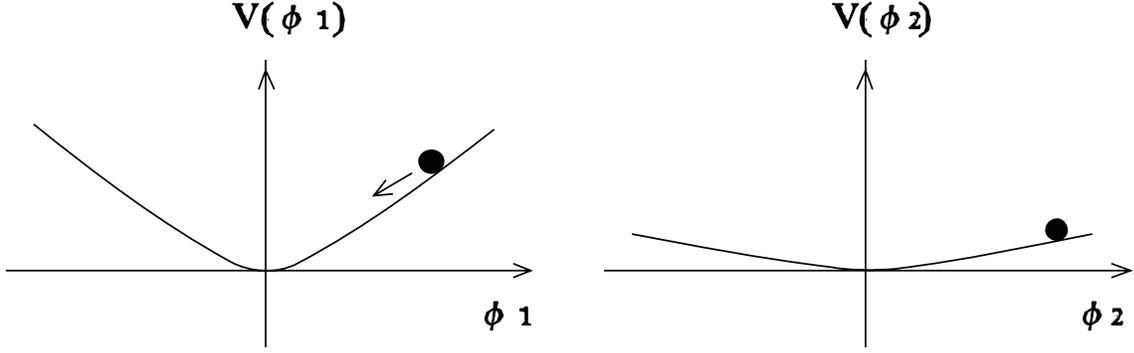}} 
\end{picture}
\caption{This figure shows the potential for badly broken symmetry
  ($x\gg 1$). $\phi_1$ oscillates to induce preheating, while $\phi_2$
  induces $m_\chi$-perturbation to the preheating.}
\label{fig3}
 \end{center}
\end{figure}

Let us explain what happens using the potential depicted in Fig.4.
Besides the hybrid curvatons we considered in this paper, one may
consider an alternative to the curvatons. 
Namely, trapped inflation is realized by the $\phi_2$-potential that
is depicted in Fig.4, which has a potential similar to thermal inflation.
Instead of considering thermal trapping, we considered in
Ref. \cite{Multi_field_trap} the trapping induced by the preheat field.
This scenario is based on the trapped inflation\cite{beauty_is} 
that has originally been discussed with string model, but the situation
discussed in Ref. \cite{Multi_field_trap} is rather different from the
original string model. 
\begin{figure}[ht]
 \begin{center}
\begin{picture}(550,120)(0,0)
\resizebox{15cm}{!}{\includegraphics{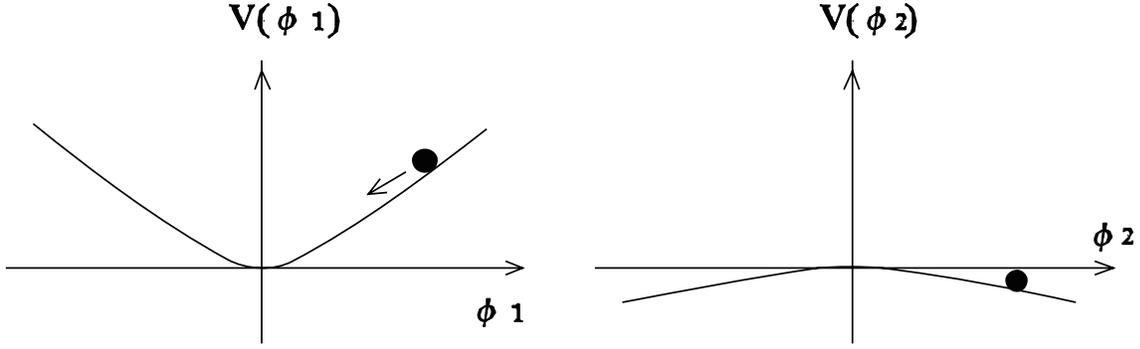}} 
\end{picture}
\caption{This figure shows the potential for a trapped inflationary
  model that has been discussed in Ref. \cite{Multi_field_trap}.
  In this case the primordial isocurvature perturbation
  $\delta \phi_2$ is converted into $\delta n_\chi$ through the
  inhomogeneous preheating, which leads to the fluctuation of the number
  of e-foldings $\delta N_e$. See Ref. \cite{alternate} for more details
of the mechanism for generating the curvature perturbation at the end of
  inflation.}
\label{fig4}
 \end{center}
\end{figure}

Finally, we will comment on the possibility that $\phi_2$ may have
quintessential potential, which is shown in Fig.5.
This model has been discussed previously by us in
Ref. \cite{NO_Curvatons}. 
We showed that the curvaton scenario is successful in this
model. 
Moreover, this model can be regarded as the hybrid of chaotic-type
inflation and quintessential inflation connected at the ESP. 

\begin{figure}[ht]
 \begin{center}
\begin{picture}(550,120)(0,0)
\resizebox{15cm}{!}{\includegraphics{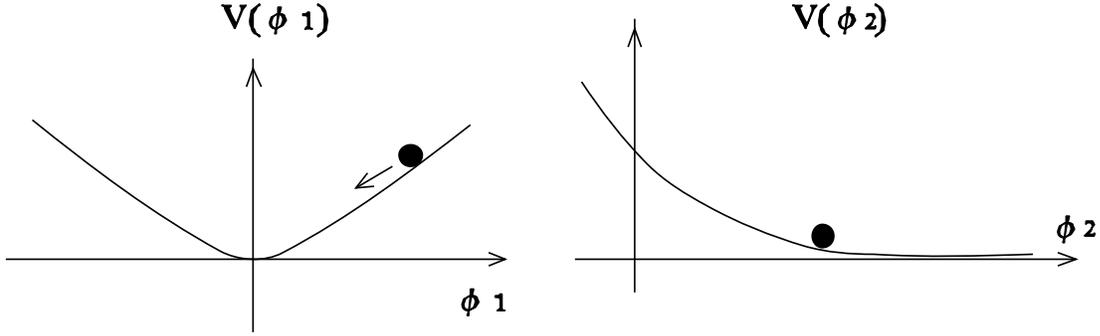}} 
\end{picture}
\caption{This picture shows the potential for hybrid-type quintessential
  inflaton that has been discussed in Ref. \cite{NO_Curvatons}.}
\label{fig5}
 \end{center}
\end{figure}
Now we would like to summarize the motivation and the results obtained
in these successive approaches. 
In these seemingly different models we considered hybrids of
inhomogeneous preheating and other cosmological scenarios (such as
curvatons, trapped and quintessential inflation) to obtain new
alternatives to the traditional inflationary scenario. 
We know that preheating is a common phenomenon that is expected to occur
at the end of inflation, or when a field starts to oscillate at a later
epoch. 
It is also very natural to expect that the potential is characterized by
broken symmetry. 
Motivated by these observations, we considered in this paper the
inhomogeneous preheating characterized by broken symmetry. 
For the instant-decay scenario, generation of the perturbations has been
discussed in Ref. \cite{SSB-inst, matsuda_inst1, matsuda_inst2}.
Our present model is different from these scenarios, as we considered in
this paper the late-decay curvatons.

\section{Acknowledgment}
We wish to thank K.Shima for encouragement, and our colleagues at
Tokyo University for their kind hospitality.
\appendix
\section{Instant-decay scenario with slightly broken symmetry}
In this appendix we will explain how the fluctuation related to the
number density of the preheat field is generated from the oscillation
using the potential characterized by slightly broken symmetry. 
Although what we will explain here has already been discussed in 
Ref. \cite{SSB-inst}, we think it is important to review the issue from
the viewpoint of applying the inhomogeneous preheating to the curvaton
generation. 
It is also important to examine the required condition in the original
model and explain how it could be removed or relaxed in the hybrid model. 

Here we will consider the potential of the scalar field described by 2
degrees of freedom ($\phi_1$ and $\phi_2$)
\begin{equation}
V(\phi_1, \phi_2) = \frac{m^2}{2}\left[ \phi_1^2 + 
\frac{\phi_2^2}{1+x}\right],
\end{equation}
where $x$ represents a measure of the symmetry breaking. 
Since we are considering slightly broken symmetry $x\ll 1$,
it would be useful to introduce a complex field $\phi$ that is
defined as
\begin{equation}
\phi=\phi_1 + \phi_2 = |\phi|e^{i\theta}.
\end{equation}
The oscillation of the scalar field $\phi$ will induce preheating for
the preheat field $\chi$ if there is interaction given by ${\cal L}=
-\frac{1}{2}g^2 |\phi|^2 \chi^2$.
In the original scenario the preheat field was assumed to have a
negligible mass term, $m_\chi^{bare} \simeq 0$.\footnote{Please note
that this is not true in our hybrid curvaton model.}
Although the bare mass is negligible, there is the effective mass
$m_\chi(\phi)$ that depends on the time-dependent value of $\phi$.
Since the preheat field becomes massless near the enhanced symmetric
point (ESP) at $|\phi|=0$, $\chi$ will be generated near the ESP. 
Then the comoving number density of the preheat field $\chi$ produced at
the first scattering is given by
\begin{equation}
n_\chi^{(1)} = \frac{(g|\dot{\phi}_*)^{3/2}}{(2\pi)^3}\exp \left[
-\frac{\pi g |\phi_*|^2}{|\dot{\phi}_*|}\right],
\end{equation}
where $*$ denotes the value when the oscillating field reaches the
minimum of the potential along its trajectory.
Expressing the initial conditions as $|\phi_0|$ and $\theta_0$,
these values are given by
\begin{eqnarray}
|\phi_*| &\simeq& \frac{|\phi_0|\pi x}{2\sqrt{2}}|\sin 2\theta_0|\\
|\dot{\phi}_*| &\simeq& m |\phi_0|\sqrt{1- x \sin^2 \theta_0}.
\end{eqnarray}
Considering the primordial perturbation $\delta \theta_0$, one obtains
\begin{eqnarray}
\frac{\delta |\phi_*|}{|\phi_*|} &\simeq& 
\frac{2\cos 2\theta_0}{\sin 2\theta_0} 
 \delta \theta_0 -(\delta \theta_0)^2\\
\frac{\delta |\dot{\phi}_*|}{|\dot{\phi}_*|} &\simeq&
-\frac{x}{2}\left[\sin 2\theta_0 \delta \theta_0 + \cos 2\theta_0 
(\delta \theta_0)^2\right].
\end{eqnarray}
Then the fluctuation that could be induced for the number density of the
preheat field is given by
\begin{eqnarray}
\frac{\delta n_\chi}{n_\chi} &\simeq &
-\frac{3x}{4}
\left[\sin 2\theta_0  \delta \theta_0 + \cos 2\theta_0 
(\delta \theta_0)^2\right].
\end{eqnarray}
where we did not consider terms proportional to $x^2$,
since they are very small in the limit $x \ll 1$.
Because of the exponential factor appearing in the form of $n_\chi$, the
condition that is needed for successful preheating is 
\begin{equation}
\label{succ_pre}
\frac{\pi g |\phi_*|^2}{ |\dot{\phi}_*|} \simeq \frac{\pi g(\phi_0 x 
\sin 2\theta_0)^2}{m\phi_0}   <1,
\end{equation}
which suggests that $x^2 < m/(g |\phi_0|\sin^2 2\theta)$.
Since in the original scenario $\phi$ is assumed to be the inflaton of a
chaotic-type inflationary model, there is an upper bound for $m$ 
($m < 10^{13}GeV$), so that the traditional perturbation generated by
the inflaton does not dominate the primordial perturbation. 
As a result, $x$ must satisfy the condition 
\begin{equation}
\label{ap1}
x^2 < 10^{-5} g^{-1} \sin^{-2} 2\theta
\end{equation}
for $\phi_0 \simeq M_p$.
Moreover, if the cosmological perturbation is generated by the
inhomogeneous preheating, there is the condition 
\begin{equation}
\label{curv-smallx}
\left[\frac{3x \sin 2\theta_0}{4}\right]\delta \theta_0 \sim 10^{-5},
\end{equation}
which leads to
\begin{equation}
\label{ap2}
x \sim 10^{-5} \frac{4}{(3 \sin 2\theta_0) \delta\theta_0}.
\end{equation}
Besides the above condition, there is another condition coming from the
non-Gaussian parameter $f_{NL} < 100$, which is 
\begin{equation}
\label{ap3}
x >0.01 \frac{\cos 2\theta_0}{ \sin^2 2\theta_0}.
\end{equation}
In order to satisfy Eq.(\ref{ap1}) and (\ref{ap3}),
the coupling constant must satisfy the condition
\begin{equation}
\label{app-con1}
g< 10^{-5}\left(\frac{\hat{f}_{NL} \sin 2\theta_0}{\cos 2\theta_0}\right)^2,
\end{equation}
where $\hat{f}_{NL}$ denotes the upper bound for $f_{NL}$.
According to the original analysis\cite{instant-original}, the ratio of
the energy density of the preheat field to the energy density of the
inflaton field will be $r_0\equiv \frac{\rho_\chi}{\rho_\phi} \sim
2g^{5/2}$, which suggests 
\begin{equation}
\label{sev_eq}
r_0 < 10^{-25/2}\left(\frac{\hat{f}_{NL} \sin
2\theta_0}{\cos 2\theta_0}\right)^{5},
\end{equation}
while $r_0 \sim 1$ is required if the cosmological perturbation is
supposed to be generated by the preheat field. 

Besides the above example which is characterized by $x \ll 1$, we
considered in Ref. \cite{matsuda_inst1, matsuda_inst2} a new model that 
is characterized by $x\gg 1$, where the lighter field $\phi_2$ is 
almost independent of the oscillating field.

Unlike the model discussed in Ref. \cite{SSB-inst, matsuda_inst1,
matsuda_inst2} in which the preheat field is assumed to decay instantly
after preheating, we will ``not'' assume in this paper that the preheating
is accompanied by instant decay. 
We will assume that the decay rate of the preheat field is rather small
so that the preheat field decays late. 
A small decay rate is due to small couplings to light fields. 
This gives a complementary scenario for the instant decay scenario. 
Assuming that $\phi$ starts to roll when $m \sim H$, the required
condition for the stability of the preheat field at the maximum value of
the oscillating field is 
\begin{equation}
g_d^2 \times g< \frac{m}{\phi_{max}},
\end{equation}
where we assumed that the decay rate is given by $\Gamma_\chi = g_d^2
\times g\phi_{max}$, where $\phi_{max}$ ($\phi_{max}\ll \phi_{osc}$) 
denotes the maximum value of the oscillating field after the first
scattering.\footnote{$\phi_{max}$
is much smaller than $\phi_{osc}$ because of the confining 
potential induced by $n_\chi\ne 0$. See Ref. \cite{beauty_is} for more
details.} 
Since the preheat field does not decay instantaneously after preheating,
it evolves like a massive matter field when it has a bare mass term. 
Since the ratio of the massive matter to the radiation energy density
increases during the radiation-dominated era, we do not have to assume
that the density of the preheat field dominates the Universe just after
preheating.   
This removes the condition $r_0 \sim 1$. 
As a result, the oscillating field that induces preheating is not
necessarily the inflaton. 
Remember that in the Affleck-dine mechanism the late-time oscillation
induced by a light scalar field plays an important role in generating
the baryon asymmetry of the Universe. 
We showed that the generation of the perturbed curvaton is possible
even if the inhomogeneous preheating is induced by late-time oscillation.

\end{document}